\documentclass[aps,prb,twocolumn,superscriptaddress]{revtex4}

\usepackage[dvipdfm]{graphicx,color}
\usepackage{bm}
\usepackage{graphicx}
\usepackage{amsmath,amssymb}
\usepackage{times}

\begin{document}

\title{Hybrid Simulation between Molecular Dynamics and Binary Collision Approximation Codes for Hydrogen injection onto  Carbon Materials}

\author{Seiki Saito}\email[]{saito.seiki@nifs.ac.jp}
\affiliation{Department of Energy Engineering and Science, Graduate school of Engineering, Nagoya University, Furo-cho, Chikusa-ku, Nagoya 464-8603, Japan.}

\author{Atsushi M. Ito}
\affiliation{National Institute for Fusion Science,322-6 Oroshi-cho, Toki 509-5292, Japan.}

\author{Arimichi Takayama}
\affiliation{National Institute for Fusion Science,322-6 Oroshi-cho, Toki 509-5292, Japan.}

\author{Takahiro Kenmotsu}
\affiliation{Faculty of Life and Medical Sciences, Doshisha University, Kyotanabe, 610-0394, Japan.}

\author{Hiroaki Nakamura}
\affiliation{Department of Energy Engineering and Science, Graduate school of Engineering, Nagoya University, Furo-cho, Chikusa-ku, Nagoya 464-8603, Japan.}
\affiliation{National Institute for Fusion Science,322-6 Oroshi-cho, Toki 509-5292, Japan.}

\begin{abstract}
Molecular dynamics (MD) simulation with modified Brenner's reactive empirical bond order (REBO) potential is a powerful tool to investigate plasma wall interaction on divertor plates in a nuclear fusion device.
However, MD simulation box's size is less than several nm for the performance of a computer.
To extend the size of the MD simulation, we develop a hybrid simulation code between MD code using REBO potential and binary collision approximation (BCA) code.
Using the BCA code instead of computing all particles with a high kinetic energy for every step in the MD simulation, considerable computation time is saved.
By demonstrating a hydrogen atom injection on a graphite by the hybrid simulation code, it is found that the hybrid simulation code works efficiently in a large simulation box. 
\end{abstract}

\maketitle

\section{Introduction}
To understand a mechanism of  chemical and physical interactions between hydrogen plasmas and  divertor plates in a nuclear fusion device, it is necessary to study an  elementary processes of the reactions.
In molecular dynamics (MD) simulations, the equations of motion of  atoms are solved numerically.
We investigated plasma-wall interaction on the divertor plates made of carbon materials by the MD simulation with modified Brenner's reactive empirical bond order (REBO) potential~\cite{cite01, cite02} in the previous works~\cite{cite03,cite06,cite07}.
A typical scale length of the MD simulation box in these works are several nm.

In order to expand the simulation box to more realistic scale length, i.e., several $\mu$m, we develop a hybrid simulation code between the MD simulation code with the REBO potential and atomic collision in any structured target(AC$\forall$T) code~\cite{cite10}.
In the AC$\forall$T code, which uses binary collision approximation (BCA), a two-body interaction is calculated instead of computing all particles for every step in the MD simulation.
Therefore computation time is saved.
We choose the AC$\forall$T code as the partner of the MD simulation for its fast calculation of a location and a velocity of the particle.
The main problem of hybridization between the two simulation codes is how to define a threshold kinetic energy of the injection particle. 
For higher kinetic energy of particle than the threshold kinetic energy, we calculate the trajectory of the particle by the AC$\forall$T code. 
As the kinetic energy is dissipated to its surroundings and attaches to the threshold kinetic energy $E_\mathrm{th}$, we give a position and a velocity of the projectile to the MD simulation code as the initial condition.
Then we perform the MD simulation.
Generally speaking, because a multi-body interaction with the neighborhood of the projectile must be calculated in the MD simulation code with the REBO potential, computation time becomes longer than in the AC$\forall$T code.
To make the performance of the simulation better, it is necessary to shorten the MD simulation and extend the AC$\forall$T simulation box.
However, it is known that the AC$\forall$T code cannot calculate exactly the particle motion in a low energy.
Comparing the MD simulation with the AC$\forall$T simulation, two simulation results are consistent in higher kinetic energy than 200 eV~\cite{cite12}.
Thus we set $E_\mathrm{th}$=200 eV at which a simulation algorithm changes from the AC$\forall$T simulation to the MD simulation.

 
\section{AC$\forall$T-MD Hybrid Simulation}\label{sec2}
A concept of hybrization between the MD simulation and the AC$\forall$T simulation is that the AC$\forall$T simulation works when a kinetic energy of a projectile is higher than the threshold energy $E_\mathrm{th}=200$ eV, while the MD simulation works when the kinetic energy becomes lower than the threshold energy $E_\mathrm{th}$.

\begin{figure}[!h]
\begin{center}
\includegraphics{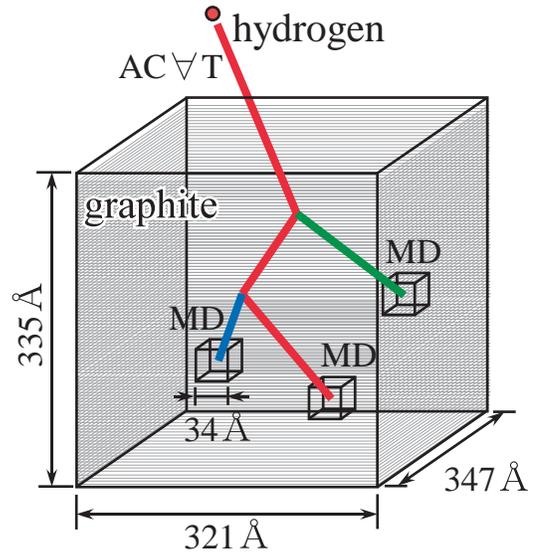}
\end{center}
\caption{Schematic diagram of the AC$\forall$T-MD hybrid simulation of hydrogen injection onto a graphite.}
\label{fig03}
\end{figure}

To explain the hybrid simulation more concretely, let's consider a hydrogen atom injection onto a graphite (Fig.~\ref{fig03}).
An injection kinetic energy of the hydrogen atom is set to 1 keV, which is higher than $E_\mathrm{th}$.
While the kinetic energy of the hydrogen atom is higher than $E_\mathrm{th}$, trajectories of the hydrogen atom and the surrounding carbon atoms are calculated by the AC$\forall$T simualtion.
The hydrogen atom dissipates its kinetic energy by interacting with carbon atoms in diamond and then the kinetic energy becomes lower than $E_\mathrm{th}$.
At that moment, the MD simulation starts to calculate the motions of the hydrogen and the carbon instead of the AC$\forall$T.
Some of carbon atoms which collide with the incident hydrogen are kicked out on the way of the AC$\forall$T simulation.
The motions of all kicked carbons in the cascade process are solved by the same procedure as the incident hydrogen.

\begin{figure}
\begin{center}
\includegraphics{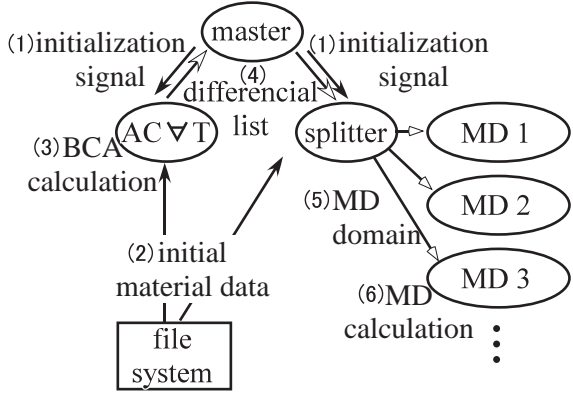}
\caption{Data flow diagram between the processes of the AC$\forall$T-MD hybrid simulation code.}
\label{fig04}
\end{center}
\end{figure}

In order to realize the above hybrid simulation, we use Multiple Program Multiple Data (MPMD)~\cite{cite11} which was developed to execute different programs parallely in one simulation with synchronizing their data.
Using the MPMD, we control four programes, i.e., (i) master program, (ii) AC$\forall$T program, (iii) MD program and (iv) splitting program as shown in Fig.~\ref{fig04}.
The master process is installed between the AC$\forall$T and MD processes for data synchronization.
After an initialization of each process (Fig.~\ref{fig04}(1) and (2)), the AC$\forall$T process (3) is executed until the kinetic energy of the incident hydrogen and all kicked atoms become lower than $E_\mathrm{th}$.
When the kinetic energies of all atoms becomes lower than $E_\mathrm{th}$, the AC$\forall$T process sends, to the master process, a list of the position and velocity of the moving atoms as a result of the AC$\forall$T simulation.
The master process (4) informs the splitter process of the completion of the AC$\forall$T process with the list of moved atoms.
The splitter process updates its material data with the list of moved atoms.
(5) Then, the MD simulation box are picked up from the entire target material and they are sent to the MD processes.
The MD simulation is started in each MD processes. (6)
We will explain the algorithm of the AC$\forall$T simulation, the MD simulation and the splitting program in the following subsections.

\subsection{AC$\forall$T Simulation}
A binary collision between a projectile and a target atom is necessary in the AC$\forall$T simulation~\cite{cite09}.
During the binary collision, a total energy, a total momentum and a total angular momentum are conserved.
Due to these conservations, the final position and velocity of the projectile and the target atom are obtained analytically on a potential $V(r)$, where $r$ is the distance between the projectile and the target atom.
In the AC$\forall$T code similarly to the ACAT code, the Moliere approximation is adopted and then $V(r)$ is given by the Thomas-Fermi potential as follows:
\begin{align}
V(r)      &= \frac{Z_1Z_2e^2}{r}\Phi(r/a), \\
\Phi(x) &= 0.35\mbox{e}^{-0.3x}+0.55\mbox{e}^{-1.2x}+0.10\mbox{e}^{-6.0x},
\end{align}
where $a$ is a screening length,  $Z_1$ and $Z_2$ are  atomic numbers of the projectile and the target atom, respectively.
Unlike the MD simulation, the AC$\forall$T simulation directry gives us the asymptotic trajectory of the projectile collided with the target.
In the AC$\forall$T simulation, the projectile hits the target atoms and then it brings about cascade shower in the large simulation box.

\subsection{Molecular Dynamics Simulation}
The equation of motion in the MD simlation is written by
\begin{equation}
\dot{{\bm r}_i} = \frac{{\bm p}_i}{m_i}, \ \ \ \dot{{\bm p}_i} = -\frac{\partial U}{\partial {\bm r}_i},
\end{equation}
where ${\bm r}_i, {\bm p}_i$ and $m_i$ are the position, momentum and the mass of the $i$-th atom. 
The function $U$ is total potential energy.
In our MD simulation, the modified REBO potential is used for the potential $U$ which is defined as:
\begin{equation}
U \equiv\sum_{i,j>i}U_{ij}, \ \ U_{ij} \equiv V_{[ij]}^{\rm R}(r_{ij}) - \bar{b}_{ij}(\{r\}, \{\theta^{\rm B}\}, \{\theta^{\rm DH}\}) V_{[ij]}^{\rm A}(r_{ij}), \label{eq4}
\end{equation}
where $r_{ij}$ is the distance between the $i$-th and the $j$-th atoms. 
The functions $V_{[ij]}^{\mathrm{R}}$ and $V_{[ij]}^{\mathrm{A}}$ represent repulsion and attraction, respectively.
The function $\bar{b}_{ij}$ generates a multi-body force.
The bond angle $\theta_{jik}^\mathrm{B}$ is the angle between the vector from the $i$-th atom to the $j$-th atom and the vector from the $i$-th atom to the $k$-th atom.
The dihedral angle $\theta_{kijl}^\mathrm{DH}$ is the angle between the plane passing through the $k$-th, $i$-th, and $j$-th atoms and the plane passing through the $i$-th, $j$-th, and $l$-th atoms. 
Moreover, the parameters $\{r\}$, $\{\theta^\mathrm{B}\}$, and $\{\theta^\mathrm{DH}\}$ denote all sets of $r_{ij}, \theta_{jik}^\mathrm{B},$ and $\theta_{kijl}^\mathrm{DH}$, respectively (for details of the modified Brenner REBO potentialm, see Refs. \cite{Ito08c} and \cite{Ito08d}).
The second-order symplectic integration\cite{Suzuki} is used to execute the time integration of the equation of motion where the time step is set to $5 \times 10^{-18} \mathrm{~s}$. 

\begin{figure}
\begin{center}
\includegraphics{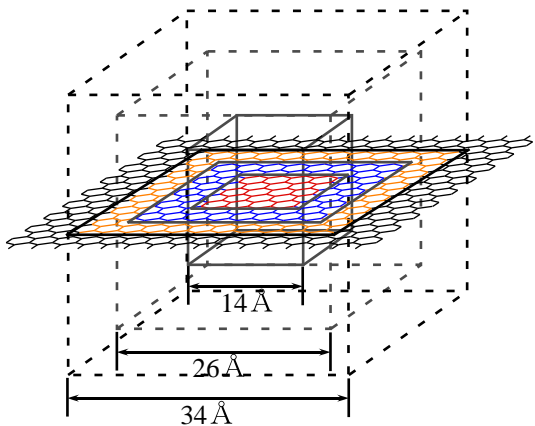}
\caption{Structure of a MD simulation box.}
\label{fig05}
\end{center}
\end{figure}

Unlike the ACAT simulation, the MD simulation cannot treat a large scale material.
To calculate efficiently, local boxs $14~\mbox{\AA}$ on a side are picked up, and the simulation for the atoms in each box is executed in parallel.
To obtain precise forces acting on all atoms in the MD simulation boxs $14~\mbox{\AA}$ on a side, the bond energy $U_{ij}$ in Eq.~(\ref{eq4}) is calculated in a larger cubic box $26~\mbox{\AA}$ on a side.
Furthermore, the function $\bar{b_{ij}}$ depends on more outer covalent bonds.
Consequently, the MD process needs the positions of atoms in a cubic box $34~\mbox{\AA}$ on a side.
Namely, in the MD process, the equations of motions are solved about atoms in the cubic box $14~\mbox{\AA}$ on a side, and forces acting on the atoms are calculated by aggregating surrounding atoms in a cubic box $34\mbox{\AA}$ on a side as shown in Fig. \ref{fig05}.

\subsection{Detamination of MD Simlation Box}
\begin{figure}
\begin{center}
\includegraphics{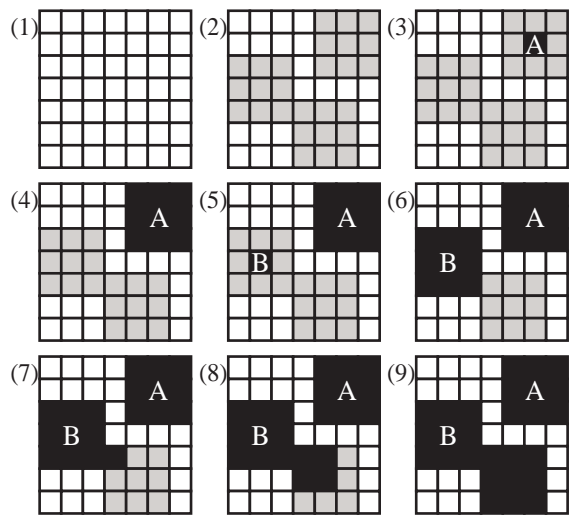}
\caption{Example of the determination procedure of the MD simulation box. The gray and white cells indicate the marked cells as MD region, respectively. The black cells indicate the counted cells by the recursion routine. (1)Dividing the target material into cell. (2)Marking the cells corresponding to three target atoms. (3),(4)Counting the MD simulation box A by recursion routine. (5)-(9)Counting the MD simulation B which includes two target atoms.}
\label{fig06}
\end{center}
\end{figure}

By the MPMD method, one CPU is allocated into the MD process treating one cubic box $34\mbox{\AA}$ on a side.
However, there is the special case that two or more cubic boxs overlap because the moving particles are close.
These overlapping boxs are regarded as a combined MD simulation box for one CPU process.
The determination for the MD simulation box is carried out by the splitting program (See Fig.~\ref{fig06}) as follows.
The simulation box in the AC$\forall$T code is divided into cubic cells $2~\mbox{\AA}$ on a side, where $2~\mbox{\AA}$ is the cutoff length of covalent bond in the REBO potential.
$17\times 17\times 17$ cells, which are corresponding to the cubic box $34~\mbox{\AA}$ on side, are marked up.
Only a center cell of the marked cells is a member of the MD simulation box, initially.
The center cell is set to the first target cell and the member of the MD simulation box is selected by the following recursive routine:
if adjoining cells of the target cell has been marked up, they are added into the member of the MD simulation box and become next target cells, while if they have not been marked up, the recursive routine is stopped.

\section{Numerical Demonstration}\label{sec3}
We demonstrate an AC$\forall$T-MD hybrid simulation for a hydrogen atom injection onto a graphite, whose size is $321~\mbox{\AA}$ long, $347~\mbox{\AA}$ wide and $335~\mbox{\AA}$ depth which consists of 100 graphene.
Periodic boundary conditions are imposed on the horizontal direction.
The initial temperature of the graphite is set to zero Kelvin.
The hydrogen atom is injected normaly onto the surface at 1 keV.

\begin{table}
\begin{center}
\caption{The average of the depth $d_\perp$ and the horizontal distance $d_\parallel$ of 900 injections. $9\%$ trajectories which go out of the target material are excepted.}\label{tbl1}
\begin{tabular}{ccccccc||c} \hline
                & $\bar{d_\perp}$\,[\AA] & $\bar{d_\parallel}$\,[\AA] \\ \hline
AC$\forall$T-MD & $127.3\pm 31.5$ & $59.1\pm 34.3$\\
AC$\forall$T    & $127.2\pm 37.5$ & $82.13\pm 45.8$\\ \hline
\end{tabular}\\
\end{center}
\end{table}

\begin{figure}
\begin{center}
\includegraphics{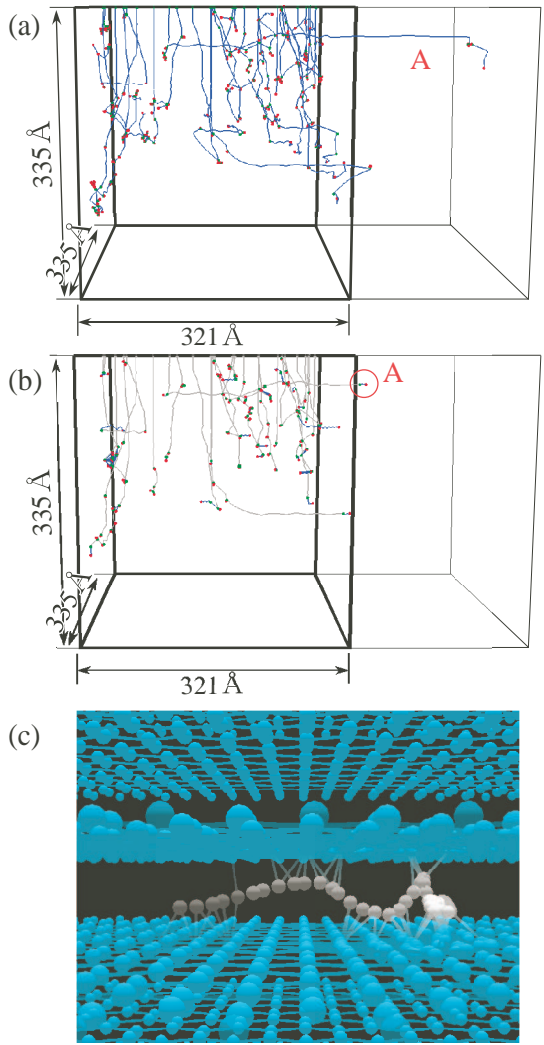}
\caption{30 sample trajectories of incident hydrogen that move in graphite calculated by the (a)stand alone AC$\forall$T code and (b)AC$\forall$T-MD hybrid simulation, respectively. (c)Detailed image of the MD simulation part for trajectory `A' in (b).}
\label{fig07}
\end{center}
\end{figure}

900 simulations with the same initial material and randomly changed injection position have been performed.
As a bench-mark, stand alone AC$\forall$T simulations have also been performed.
Table~\ref{tbl1} shows the average and the standard deviation of the depth $d_\perp$ and the horizontal distance $d_\parallel$ from the surface to the final positions in the result of both the AC$\forall$T-MD hybrid simulation and the AC$\forall$T simulation.
Although the averages of depth in the both cases are almost equal, the average and the standard deviation of horizontal distance by the AC$\forall$T-MD hybrid simulation are smaller than that of the AC$\forall$T simulation.
Figure \ref{fig07}(a) and (b) show 30 samples chosen randomly in 900 trajectories calculated by the AC$\forall$T and the AC$\forall$T-MD hybrid code, respectively.
Figure \ref{fig07}(b) shows the trajectories in the same initial condition calculated by the AC$\forall$T-MD hybrid simulation.
The gray and blue lines in Fig.~\ref{fig07}(b) show the trajectories of which indicate the AC$\forall$T contribution and the MD contribution, respectively.
Comparing these two figures, it is found that the AC$\forall$T-MD hybrid simulation gives shorter trajectories than the AC$\forall$T simulation does.
As shown in the trajectory `A' in Fig.~\ref{fig07}(a), the incident hydrogen atom moved the long distance along the interlayer region of the graphite, because the AC$\forall$T code does not treat the binding from surrounding atoms.
The MD simulation includes it by the multi-body force.
Therefore, the atom is not possible to move a long distance for low kinetic energy in the AC$\forall$T-MD hybrid simulation as shown in Fig.~\ref{fig07}(c).
In our previous research~\cite{cite07}, it was shown that there is an adsorption site $1.1~\mbox{\AA}$ above each of the carbon on a graphene sheet.
The hydrogen calculated by the MD dissipates its kinetic energy quicker than the AC$\forall$T case and finally be trapped at the adsorption site.

\section{Summary}\label{sec4}
The AC$\forall$T-MD hybrid simulation code has been developed to treat large materials, long time and wide range kinetic energy in PSI.
The motion of hydrogen atom is solved by BCA method using the AC$\forall$T code while its kinetic energy is higher than $200~\mbox{eV}$.
When the kinetic energy becomes lower than $200~\mbox{eV}$, the MD code is invoked around atoms moved by the AC$\forall$T simulation.

The AC$\forall$T-MD hybrid simulation was performed and compared with stand alone AC$\forall$T simulations of 900 hydrogen atom injections onto a graphite.
Their results suggested that the AC$\forall$T-MD hybrid simulation tends to derives a shorter trajectory than the AC$\forall$T simulation does.
This fact is caused by the binding from the surrounding atoms, which is ignored in the AC$\forall$T simulation.


\vspace{1em}\section{Acknowledgement}\vspace{1em}
Our deep appreciation should go to Prof. Takahiro Kenmotsu of Faculty of Life and Medical Sciences, Doshisha University.
Numerical simulations were carried out using the Plasma Simulator at the National Institute for Fusion Science, Japan.
This work is supported by the National Institutes of Natural Sciences' undertaking Forming Bases for Interdisciplinary and International Research through Cooperation Across Fields of Study and Collaborative Research Program (No. NIFS09KEIN0091) and Grants-in-Aid for Scientific Research (No. 19055005) from the Ministry of Education, Culture, Sports, Science and Technology, Japan.
This work is performed with the support and under the auspices of the NIFS Collaboration Research program (NIFS09KTAN008). 


\end{document}